\documentclass[epjc3,final]{svjour3}

\RequirePackage{graphicx}

\journalname{Eur. Phys. J. C}

\usepackage{amsmath}
\usepackage{graphicx}
\usepackage{pifont}
\usepackage{array}
\usepackage{rotate}
\usepackage{vmargin}
\usepackage{fancyhdr}
\usepackage{color}
\usepackage{lineno}
\usepackage{url}
\usepackage{multirow}
\usepackage{wasysym}

\newcommand{\lapprox}{_<\atop^\sim}  
\newcommand{\gapprox}{_>\atop^\sim}  

\begin{document}


\sloppy


\title{Searching for magnetic monopoles trapped in accelerator material at the Large Hadron Collider}

\author{
        M.~Dam Joergensen\thanksref{addr9}
        \and
        A.~De~Roeck\thanksref{addr5,addr6,addr7}
        \and
        H.-P.~H\"achler\thanksref{addr10}
        \and
        A.~Hirt\thanksref{addr10}
        \and
        A.~Katre\thanksref{addr1}
        \and
        P.~Mermod\thanksref{e1,addr1}
        \and
        D.~Milstead\thanksref{e1,addr3}
        \and
        T.~Sloan\thanksref{addr4}
}

\thankstext{e1}{Corresponding authors. E-mail addresses are: philippe.mermod@cern.ch, milstead@physto.se.}
\institute{
           Niels Bohr Institute, Denmark\label{addr9}
           \and
           CERN, Geneva, Switzerland \label{addr5}
           \and
           Department of Physics, University of Antwerp, Belgium \label{addr6}
           \and
           Department of Physics, UC-Davis, USA \label{addr7}
           \and
           Department of Earth Sciences, Swiss Federal Institute of Technology (ETH), Switzerland\label{addr10}
           \and
           D\'epartement de Physique Nucl\'eaire et Corpuculaire, University of Geneva, Switzerland\label{addr1}
           \and
           Fysikum, Stockholm University, Sweden \label{addr3}
           \and
           Department of Physics, Lancaster University, UK \label{addr4}
}

\date{\today}

\maketitle

\begin{abstract}
If produced in high energy particle collisions at the LHC, magnetic monopoles could stop in material surrounding the interaction points. Obsolete parts of the beam pipe near the CMS interaction region, which were exposed to the products of $pp$ and heavy ion collisions, were analysed using a SQUID-based magnetometer. The purpose of this work is to quantify the performance of the magnetometer in the context of a monopole search using a small set of samples of accelerator material ahead of the 2013 shutdown.


\end{abstract}

\section{Introduction}

The discovery of particles possessing magnetic charge would be of fundamental significance. The Maxwell equations would perforce be symmetrised and models which unify the fundamental forces would receive direct experimental support~\cite{Hooft1974,Preskill1984}. Furthermore, Dirac demonstrated that the quantisation of electric charge could be understood as a consequence of the quantisation of angular momentum should monopoles exist~\cite{Dirac1931,Dirac1948}. Searches for monopoles are thus well motivated by arguments from classical physics, quantum mechanics and quantum field theory.  A wide range of searches has therefore been performed in a number of environments such as cosmic ray facilities and colliders~\cite{Fairbairn:2006gg,Milton:2006cp,PDG2010}. This paper describes first measurements of accelerator material for a search for monopoles produced at the CERN Large Hadron Collider (LHC).

 Two important quantities characterising any collider monopole search are the mass and charge range to which the search is sensitive. The mass sensitivity is largely determined by the collision centre-of-mass energy whereas the charge range depends mainly on the experimental technique used in the search. The Dirac argument suggests a value for the lowest amount of magnetic charge possessed by a particle. This is the so-called Dirac charge $g_D=\frac{nh}{e{\mu}_0}=3.3\times 10^{-9}$~Am where $e$ is the elementary electric charge, $h$ is Planck's constant, $\mu_0$ is the permeability of free space, and the quantum number $n$ is set to unity.  A modification of the Dirac argument can be made by allowing a down-type quark to carry the fundamental electric charge in which case the minimum magnetic charge becomes $3g_D$. Furthermore, by considering an object with magnetic and electric charge (a so-called dyon\footnote{The term dyon is used in this paper only in contexts in which the electric charge possessed by a magnetically charged object is relevant. Otherwise the term monopole should be understood as referring to both monopoles and dyons.}) in place of a monopole,  Schwinger argued that $n$ ought to be even~\cite{Schwinger:1966nj,Zwanziger:1969by}. The minimum magnetic charge may well therefore be as high as $6g_D$. Monopoles with charges around the Dirac charge would suffer electromagnetic energy loss in matter several thousand times that of a particle possessing the elementary electric charge~\cite{Ahlen78,Ahlen80}. Given this extreme energy loss, monopoles of different charges may either dominantly be stopped in detector and accelerator material or propagate through the detector and be observed as a highly ionising object.  Complementary searches which aim to provide a comprehensive coverage of magnetic and electric charge are therefore needed. This work forms part of a program of searches for highly charged objects at the LHC~\cite{Moedal,Pinfold:2009zz,:1999fr,Aad:2011mb,ATLASmonoconf}. Here, the focus is on stopped monopoles which correspond broadly to scenarios of monopoles with with charges $\gapprox$ $g_D$ and dyons with high electric charge ($\gapprox$ $100e$) and modest magnetic charge ($\lapprox$ $g_D$)~\cite{SMPpheno11}. Such particles would be expected to be bound strongly to matter~\cite{Milton:2006cp}.

The so-called induction method~\cite{Alvarez:1970zu,Alvarez:1971zt} has been used for searches at lower energy colliders~\cite{TEVATRONSQUID2000,TEVATRONSQUID2004,Aktas:2004qd}  and is employed here. This involves passing material samples through superconducting loops and looking for an induced non-decaying current from a transported monopole. SQUID-based magnetometers in principle offer the required precision with which to measure such a current although studies must be performed to quantify the expected response of the magnetometer to a monopole. In this work a variety of calibration tools are used together with magnetic field simulations to characterise the capability of a SQUID-based magnetometer to detect monopoles. Special attention is also paid to the problem of backgrounds from large dipole moments which can induce monopole-like residual currents. The search methodology is then tested with a small set of material samples from the LHC accelerator. The samples have been exposed to proton-proton collisions at centre-of-mass energies up to 7~TeV and lead-lead collisions at nucleon-nucleon centre-of-mass energies up to 2.76~TeV. The corresponding integrated luminosity values were around 6~fb$^{-1}$ ($pp$) and 170~$\mu$b$^{-1}$ (Pb-Pb). The aim of this work is to develop and test the experimental techniques needed for a wider study of accelerator and detector material which could become available following the shutdown and upgrade of the LHC in 2013. In particular, it is important to demonstrate that monopoles with charges of values much less and much greater than the Dirac charge could be observed.

This paper is organised as follows. The magnetometer used in this work is described and its sensitivity to objects possessing a range of magnetic charges assessed using different calibration methods. Sources of fake monopole signals are then studied with rock samples. The accelerator samples are then described together with estimations of energy loss a monopole would suffer before becoming embedded in these samples. Finally, the response of the magnetometer to the samples is shown and the results discussed.

\section{Experimental apparatus and tools}\label{sec:expmet}

 For this work, a 2G Enterprises DC-SQUID rock magnetometer (model 755)~\cite{2g} was used. The device is housed at a laboratory maintained by the Earth and Planetary Magnetism group at the Department of Earth Sciences at the Swiss Federal Institute of Technology (ETH), Zurich. The relevant apparatus within the magnetometer for this study is a flux sensing system comprising two pick-up coils of radius 4~cm along the longitudinal $z$-axis of the magnetometer.
  Samples are transported along the axis in the $+z$ direction through an access shaft with diameter $\sim 4$~cm. The sensing region is surrounded by superconducting shielding. The magnetic flux from a sample in the pick-up coils is sensed as a superconducting current in the coils.

\begin{figure}[tb]
  \begin{center}
    \includegraphics[width=0.6\linewidth,angle=270]{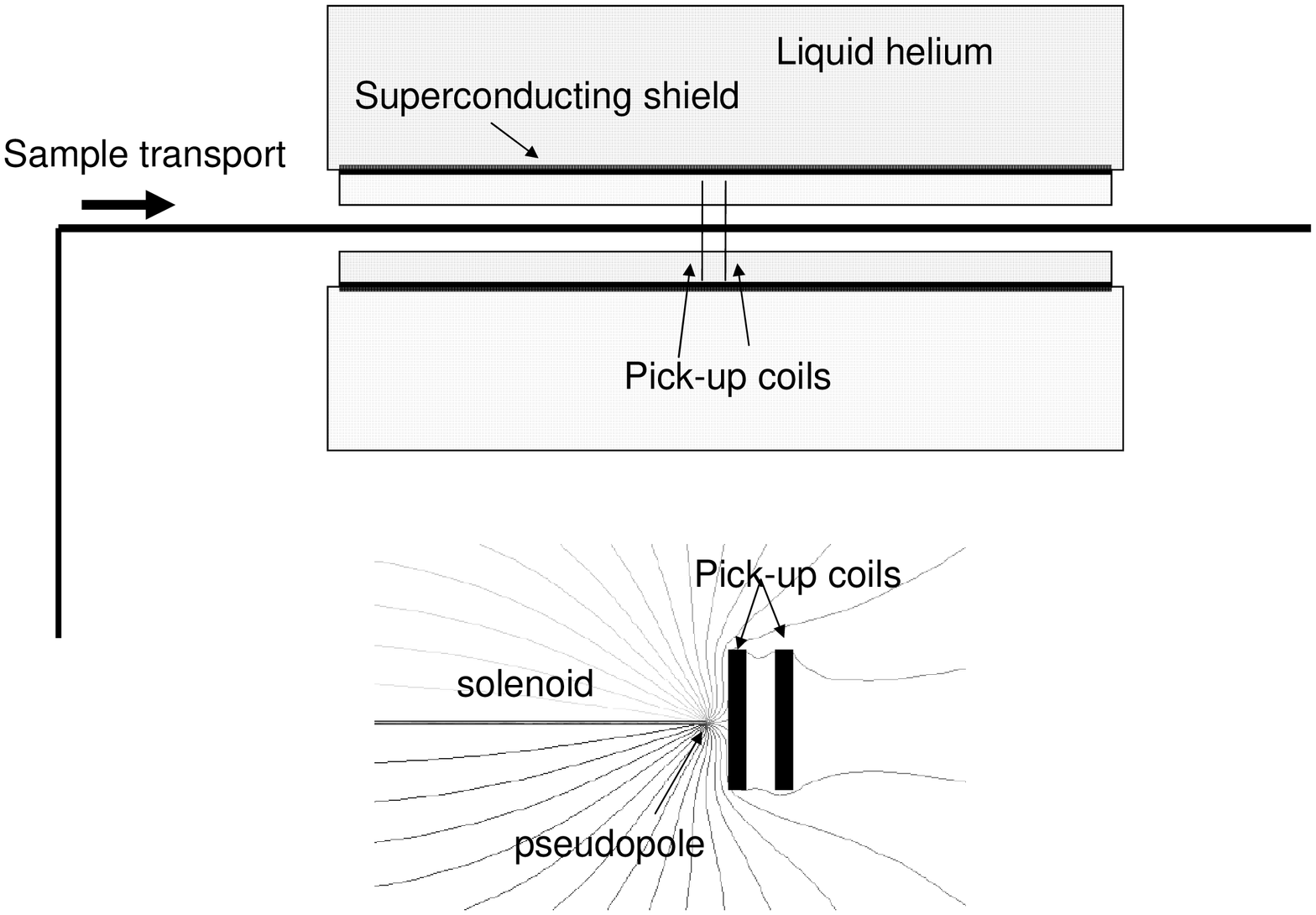}
 \end{center}
 \vspace{0.0cm}
  \caption{Top: schematic representation of the magnetometer used in this work. Bottom: magnetic field configuration of of pseudopole near to two superconducting pick-up coils. }
  \label{fig:spic}
\end{figure}

 Figure~\ref{fig:spic} shows a schematic outline of the magnetometer  together with an illustration of the magnetic field configuration arising from a solenoid, one end of which is a pseudopole, near to two superconducting pick-up coils. The field calculations were made with the Maxwell program~\cite{Maxwell}. As can be seen, the net flux through the coils is zero owing to induced screening currents which cancel the flux from the pseudopole. Exact first principle calculations of the response function of the 2G magnetometer are beyond the scope of this work since they require the precise modelling of the disposition of superconducting shielding and the magnetic field in the sensing region~\cite{parker,jackson}.

%

A so-called calibration sample in the shape of a needle of length $14$~mm and diameter $1$~mm is used for calibration purposes. This is enclosed within a non-magnetic plastic holder. The dipole sample was made from floppy disk material and subsequently magnetised such that the dipole moment, aligned along the longitudinal direction, is $3.02\times 10^{-6}$Am$^2$. The uncertainty on the moment, assessed by comparing measurements on independent magnetometers at the ETH laboratory, is less than 1\%. As described in Section~\ref{sec:calib}, measurements of the calibration sample can, with the aid of a convolution method, be used to predict the magnetometer response to a monopole.

A more direct approach to inferring the magnetometer response to a monopole is to use a long solenoid since the magnetic field from one end of an semi-infinite solenoid follows an inverse square law~\cite{chen}. For this work, two thin solenoids (solenoids 1 and 2), wound with copper wire and supported by cylindrical copper formers, were used. Solenoid 1 (2) is formed from two (three) layers of wound copper wire.  These devices were previously used for a search at the H1 experiment~\cite{Aktas:2004qd}. A solenoid with $n$ turns, length $l$, surface area $S$ and applied current $I$ can be considered as possessing two oppositely charged poles (termed pseudopoles) of strength $q=I\cdot S\cdot n/l$. Expressing the pseudopole strength in units of the Dirac charge $g_D$ and the current flowing through the solenoids, the solenoids are characterised by $32.4g_D$ and $41.4g_D$ per  unit $\mu$A, respectively. The solenoid parameters are described in Tab.~\ref{tab:solenoids} and the performance of the solenoids is described in Section~\ref{sec:calib}.

A fake monopole signal can arise when a highly magnetised material passes through the pick-up coils. In this situation, the measured current does not return to zero when the material is far from the pick-up coils, thus mimicking the residual current left by monopole. To study this effect, ferromagnetic rock samples with dipole moments of values similar to those of the accelerator samples were used. This is described in Section~\ref{sec:fake}.

\begin{table}
\centering
\begin{tabular}{|c|c|c|}
\hline
Calibration coil & 1 & 2\\
\hline
Pseudopole strength/current ($g_D$/$\mu A$) & 32.4 & 41.4\\
Coil length $l$ (mm) & $250$ & $250$ \\
Number of turns $n$ & $2750$ & $7500$ \\
Wire diameter (mm) & $0.18$ & $0.1$ \\
Number of wire layers & $2$ & $3$ \\
Mean coil area $S$ ($\mathrm{mm}^2$) & $9.7$ & $4.5$ \\
Uncertainty in area & $6\%$ & $10\%$\\
\hline
\end{tabular}
\caption{Description of the calibration solenoids}
\label{tab:solenoids}
\end{table}



\section{The expected response of the magnetometer to monopoles}\label{sec:calib}

\begin{figure}[tb]
  \begin{center}
    \includegraphics[width=0.8\linewidth,angle=0]{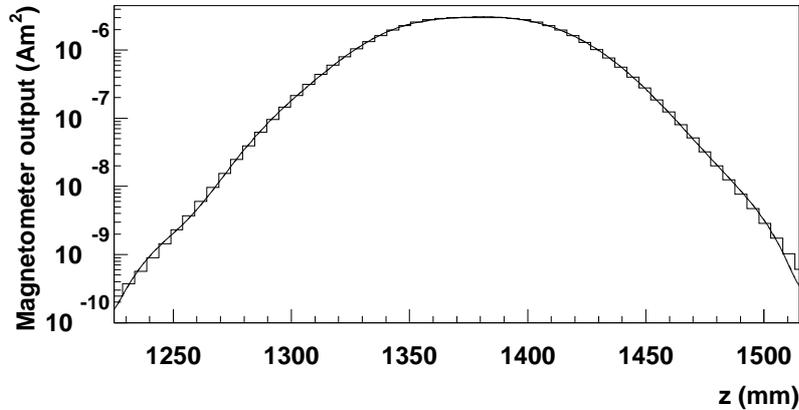}
  \end{center}
  \vspace{-6.0cm}
  \caption{The measured current from the calibration sample as a function of $z$. A smoothed form of the spectrum is overlaid. The data are expressed in units of magnetic moment since the magnetometer calibration is such that the plateau value returns the value of the sample dipole moment. }
  \label{fig:sample}
\end{figure}

\begin{figure}[tb]
  \begin{center}
    \includegraphics[width=0.8\linewidth]{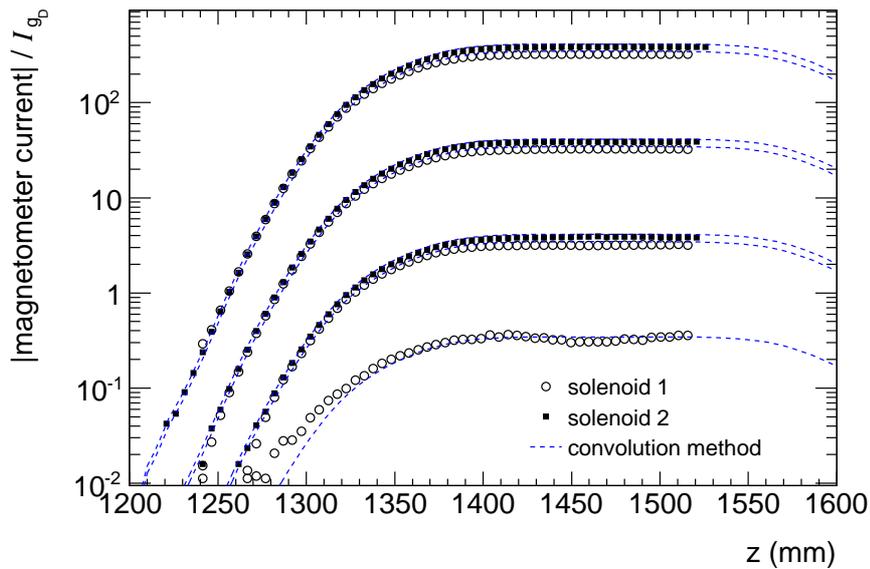}
  \end{center}
  \caption{The dependence on $z$ of the induced current for solenoids carrying a range of different currents. Also shown is the prediction of the convolution method (dashed lines) for a given solenoid carrying a certain value of current. Where two lines and two sets of data points are shown close together the upper line corresponds to the upper set of data points and {\it vice versa}.  The unit of the $y$-axis is the current expected for a Dirac monopole ($I_{g_D}$), as estimated with the convolution method.}
  \label{fig:solenoids}
\end{figure}

As mentioned in Section~\ref{sec:expmet} two approaches are employed to quantify the expected response of the magnetometer to a monopole. In one approach, the convolution method,  measurements of the current due to the calibration sample are used. Such measurements are shown as a function of the longitudinal position $z$ of the sample in Fig.~\ref{fig:sample}. The measured current rises, reaches a plateau of length roughly that of the longitudinal extent of the pick-up coil array ($\sim 4$~cm), and falls again. It should be noted that, although the device measures an induced current due to a change in magnetic field, normalisation and calibration steps have been performed such that the value of the plateau region corresponds to the magnetic moment of the sample. The distribution is thus labelled accordingly.

The superposition principle for magnetic fields implies that the field from a long thin magnet is equivalent to that of the sum of individual dipole samples positioned alongside each other such that the total length would be that of the long magnet. The distribution in Fig.~\ref{fig:sample} can therefore be used to predict the response of the magnetometer to a pseudopole. To implement this approach, the distribution in Fig.~\ref{fig:sample} was smoothed with a spline algorithm and the magnetometer response at 18 dipole positions, each separated by 14~mm, was extracted and summed. This then corresponds to a long thin bar magnet possessing a pseudopole value of $\sim 6500g_D$. The sum is therefore scaled such that it corresponds to the pseudopole charges associated with the solenoids for a range of currents, as derived using the solenoid properties in Tab.~\ref{tab:solenoids}. This procedure was performed for a ``chained" set of calibration samples at many different values in $z$.

The predicted induced current is shown (dashed lines) in Fig.~\ref{fig:solenoids} for solenoids carrying different currents. As would be expected the predicted induced current rises and reaches a plateau and begins to fall. This corresponds to the ``chained" set of samples moving towards, into and out of the pick-up coil array. The normalisation of these distributions is such that the plateau value corresponds to the strength of the magnetic pseudopole at the end of the ``chain", which is directly proportional to the current in a solenoid. The distributions is expressed in units of $I_{g_D}$, where $I_{g_D}$ is the predicted induced current from a ``chain" carrying the Dirac charge at the centre of the centre of plateau region ($z$=$1470$~mm).  The assumed current values were:  0.01, 0.1, 1 and 10 $\mu A$ for solenoid 2; the solenoid 1 current values were 7\% higher than those used in the second solenoid\footnote{The difference in current values arose due to the use of different batteries for the various solenoid runs.}.

Also shown in Fig.~\ref{fig:solenoids} are the measurements of induced current from the different solenoids with the aforementioned solenoid currents. Runs were performed in which the solenoids were stepped through the magnetometer. For each solenoid, runs were also taken with zero solenoid current. Data from runs with zero current were subtracted from the finite current runs to give the distributions in Fig.~\ref{fig:solenoids}. Solenoid measurements with a positive pseudopole entering the sensing region give positive output values and {\it vice versa}, and it was verified that there is no significant magnetic charge asymmetry in the magnetometer response. The output is found to be linear over the studied range $0.3g_D$ to $400g_D$. Good agreement is observed over several orders of magnitude of current between the induced currents left by the solenoids and those predicted by the convolution method although there is a tendency for the convolution method to slightly overestimate the solenoid measurements. Uncertainties on the solenoid measurements are not shown and are discussed below. The convolution method makes predictions at higher values of $z$ than are possible with the solenoids owing to hardware limitations in the sample transport system. A small discrepancy is evident at low values of $z$ for the lowest value pseudopole charge ($\sim 0.3g_D$) corresponding to a weak signal at which the magnetometer accuracy is degraded.

\begin{figure}[tb]
  \begin{center}
  \hspace{-0.0cm}
    \includegraphics[width=0.8\linewidth]{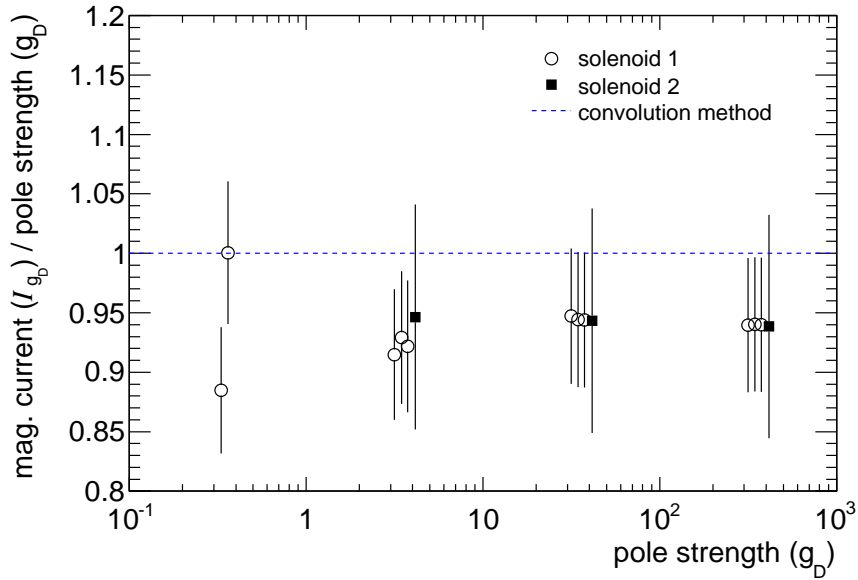}
  \end{center}

  \caption{The ratio of magnetometer current/pseudopole strength as a function of pseudopole strength for the two solenoids. The magnetometer current is the value in the plateau region in units.  The uncertainties on each point for a given solenoid are 100\% correlated. Pseudopole strength values of solenoid 1 at specific values of pseudopole strength are offset with respect to each other for clarity. In ascending order of pseudopole strength the clusters of points for solenoid 1 represent charges at $0.346g_D$, $3.46g_D$, $34.6g_D$ and $346g_D$.}
  \label{fig:solenoids2}
\end{figure}

Fig.~\ref{fig:solenoids2} allows a more quantitative comparison of the two methods. This figure shows the ratio of induced current at the centre of the plateau region to the pseudopole strength as a function of pseudopole strength. The pseudopole strengh is calculated using the solenoid parameters and solenoid current.
By construction, the ratio is unity for the convolution method. The relative uncertainties on the areas (and therefore pseudopole strength) of solenoid 1 (6\%) and 2 (10\%) are also shown. The errors on the data points in Fig.~\ref{fig:solenoids2} for a given solenoid are thus 100\% correlated with each other. As can be seen the magnetometer measurements of each solenoid are consistent with each other and the convolution method. The agreement between the convolution method and the solenoid approach is between $\sim 1$ and $\sim 1.2$$\sigma$ for solenoid 1 and is around  0.6$\sigma$ for solenoid 2. At low values of magnetic charge (and thus magnetic field) $\sim 0.3$$g_D$ larger fluctuations occur than are seen for the higher charge measurements.

A detailed error analysis for the convolution method is not performed here. However, we expect the uncertainties to be relatively small. Previous studies with with 2G SQUID magnetometers of the difference in magnetometer output for chained and long core samples gave differences of up to $\sim 10$\%~\cite{weeks}. The sample dipole moment is known to an accuracy of better than 1\%. Repeat measurements of the calibration sample lead to differences of less than 1\% when the sample is in the sensing region and the magnetic field in this region is highest. These differences can rise up to 10\% for the highest and lowest values of $z$ although these make negligible contributions to the prediction of the critical plateau value and thus expected monopole response. The convolution method relies on a number of assumptions. One such assumption is that the magnetometer output for a magnetic field strength corresponding to optimal magnetometer performance can be linearly scaled such that it predicts the performance of the magnetometer at significantly lower field values. A further assumption is that the induced current does not have a strong dependenc on sample position which could arise due to a position-dependent misalignment of the sample. Figs.~\ref{fig:solenoids} and ~\ref{fig:solenoids2} demonstrate that these assumptions are well founded.

Independently of experimental errors, differences between the methods can also arise due to the effects of the different cross sectional areas of the calibration sample and the solenoids. The Maxwell simulation program predicts that such differences are at most several per cent. Given this, and the discussion in this Section, it can be stated the calibration of the magnetometer for the passage of a monopole with charge values between $\sim 0.3g_D$ to $\sim 400g_D$ is well determined by two independent methods to an accuracy in reconstructed charge of around $10\%$. It should, however, be emphasised that this represents an ideal performance obtained with calibration tools. The next Section outlines how the performance can potentially be degraded due to machine effects and highly magnetised samples.


\section{Fake signals}\label{sec:fake}

To search for a monopole signal it is not necessary to make time-consuming repeated positions in different steps in $z$. Instead, a straightforward signal observable the {\it persistent current} is defined. This is the change in current on the pick-up coils which occurs when a sample enters the magnetometer and is transported through the pick up coils and along the axis until the magnetic field in the sensing region arising from the sample is negligible.  Two instrumental effects have been studied which may induce a fake non-zero persistent current: offset drifts and offset jumps. A third possible effect, so-called flux jumps~\cite{Levy}, did not effect this study. To investigate the first two effects, two rock samples, which were assumed not to contain monopoles, were used~\cite{R1andR2}.

 Offset drifts were observed to cause the persistent current to fluctuate with time, typically at a rate equivalent to $0.1g_D$ per hour. Since the offset drift affects the readings at all positions, the quantity defined as the last reading minus the first reading is stable with time. Using this quantity, for the background subtraction it is not necessary to make frequent empty holder passes in between sample passes since all empty holder measurements yield consistent values. The last reading minus first reading after subtracting this unique value obtained from the empty holder measurements can thus be regarded as the persistent current, the magnitude of which would be directly proportional to the candidate monopole charge.

\begin{table}
\centering
\begin{tabular}{|c|c|c|}
\hline
& Positive magnetisation & Negative magnetisation\\
\hline
\hline
Normal mode & $\mu = 0.024 \pm 0.005$ & $\mu = -0.038 \pm 0.008$\\
& $\sigma = 0.048 \pm 0.004$ & $\sigma = 0.053 \pm 0.006$ \\
\hline
Abnormal mode & $\mu = -0.143 \pm 0.004$ & $\mu = 0.005 \pm 0.007$\\
& $\sigma = 0.050 \pm 0.003$ & $\sigma = 0.059 \pm 0.005$ \\
\hline
\end{tabular}
\caption{Mean and standard deviations (in units of $I_{gD}$ ) of the persistent current distributions in various conditions using samples with magnetisation similar to the one of the beam-pipe sample. Each entry is based on $45$ to $160$ repeated measurements.}
\label{tab:fakes}
\end{table}

 Offset jumps are sudden shifts of the readings from one position to the other. Spurious offset jumps can potentially fake large signals. Such fake signals are easy to dismiss as they would not appear consistently in multiple passes. However, in certain conditions, small offset jumps were also observed to happen consistently at every pass. To better quantify this effect, we made a series of passes using rock samples. The first rock sample, R1, was chosen such that it's magnetic moment ($4.26 \times 10^{-5}$Am$^2$) is similar to those of the beampipe samples. This sample was passed multiple times into the magnetometer with orientations yielding both positive and negative values of the longitudinal magnetisation. Such measurements yielded distributions of the persistent current with mean values very near zero and very small standard deviation, as reported in the top row of Table~\ref{tab:fakes}. Significant offset jumps were observed after using another rock sample, R2, with higher magnetic moment ($1.3\times 10^{-4}$ Am$^2$), for which measurements of the persistent current in the case of a positive longitudinal magnetisation were consistently observed to be shifted by $-0.14g_D$ relatively to empty holder measurements performed in between. This shift was not observed when the sample was oriented so as to yield a negative longitudinal magnetisation. New series of measurements were then performed after demagnetisation of the R2 sample to give a moment of $1.5\times 10^{-5}$ Am$^2$ and then further to $4.5\times 10^{-6}$ Am$^2$. The same shift was observed also for the demagnetised sample, as well as for the R1 sample when measured again. It was concluded that the magnetometer has entered an abnormal mode in which any sample with positive longitudinal magnetisation, even small, systematically provokes a offset jump of magnitude $-0.14g_D$. The bottom row of Table~\ref{tab:fakes} shows the mean and standard deviation of the persistent current distribution in abnormal mode. This abnormal mode was triggered by the first pass with R2 with a high longitudinal magnetisation. It was found that the magnetometer could be restored into a normal mode by manually resetting the current offset to a value near zero. After resetting, the persistent currents for R1 and (demagnetised) R2 were again entirely consistent with the values in the top row of Table~\ref{tab:fakes}.

The resolution of the magnetometer for highly magnetised samples is dominated by offset jumps and can be considered as being the typical deviation of the persistent current from zero for samples where there is no monopole. Using the repeated rock sample measurements, it is found that in normal mode, the resolution is $0.04g_D$, and in the case where the magnetometer enters an abnormal mode, it is at least as large as $0.14g_D$.


\section{Beampipe samples}

\begin{figure}[tb]
  \begin{center}
    \includegraphics[width=1.\linewidth,angle=0]{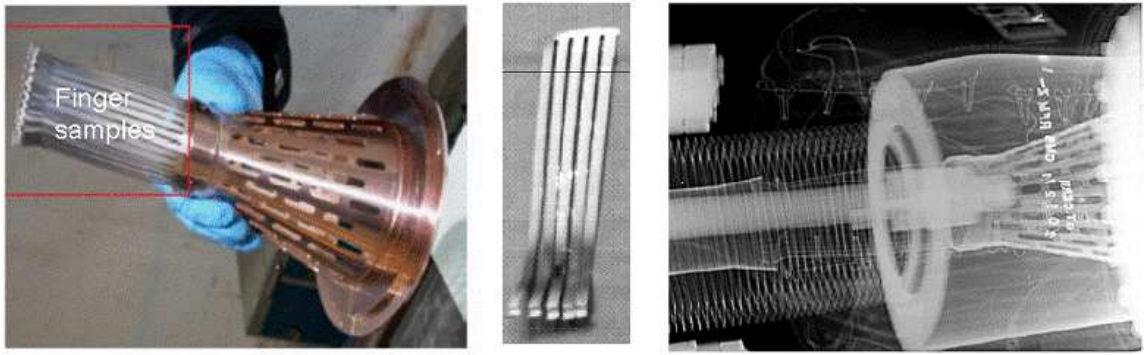}
 \end{center}
  \caption{Left: a photograph of the plugin module showing the finger samples. Middle: a photograph of one of the finger samples. Right: an X-ray picture of the plugin module attached to the CMS beampipe. }
  \label{fig:xray}
\end{figure}

\begin{figure}[tb]
  \begin{center}
    \includegraphics[width=0.45\linewidth]{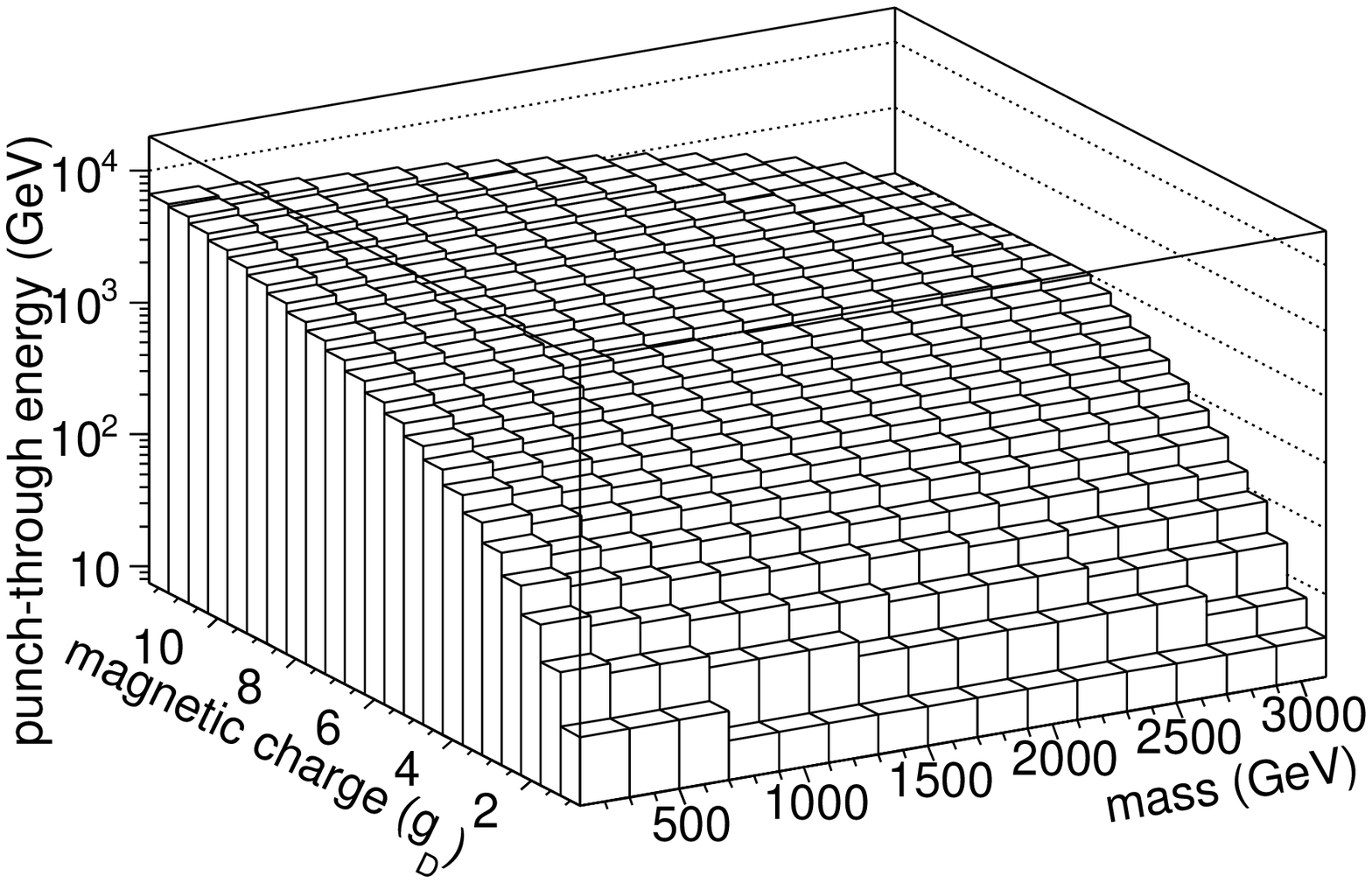}
    \includegraphics[width=0.45\linewidth]{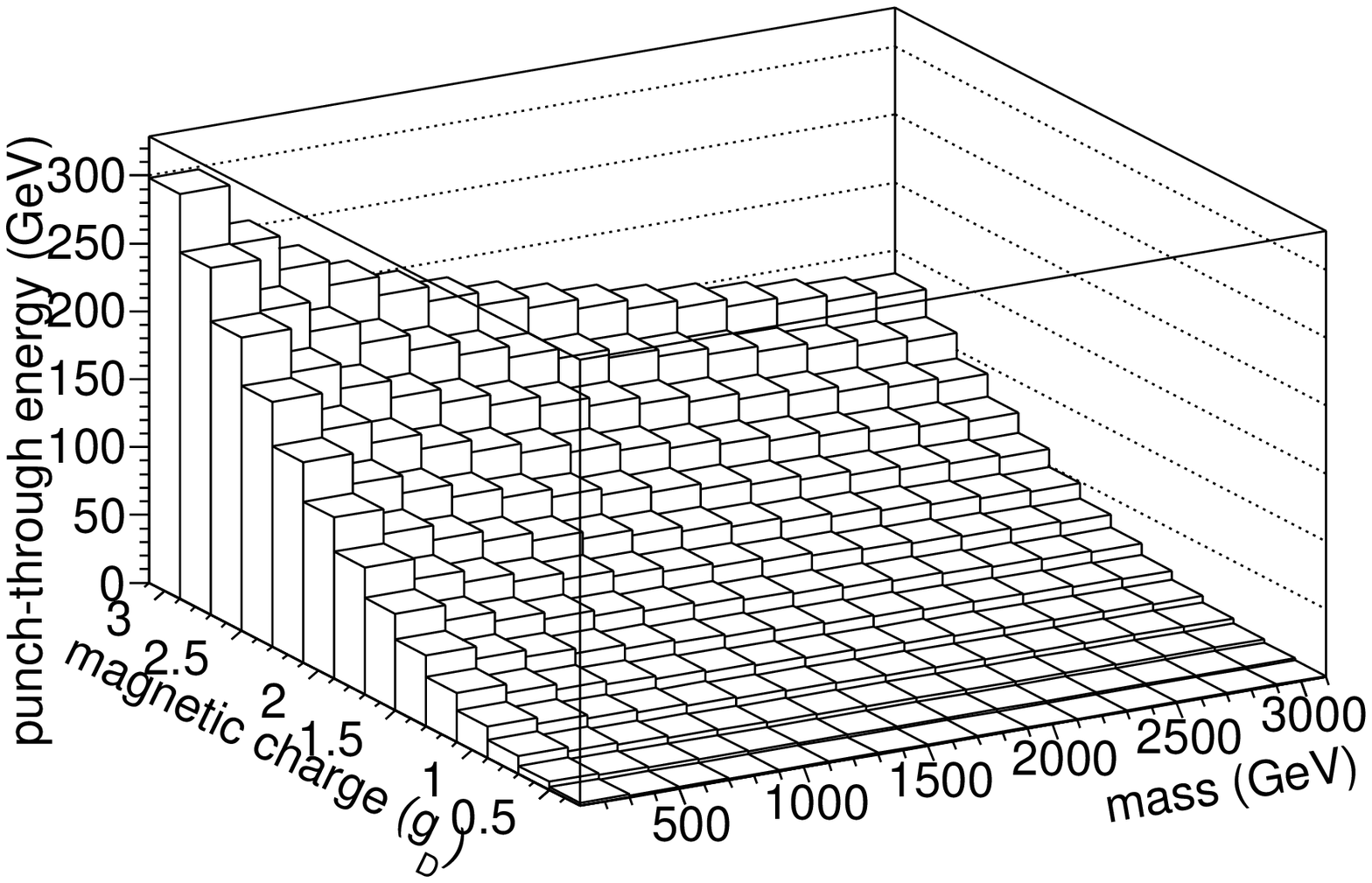}
  \end{center}
  \caption{Kinetic energy below which a monopole incident upon the beampipe sample would stop inside the sample, as a function of mass and magnetic charge. The plot on the right shows only the charges below $3g_D$ with a better precision and with a linear vertical scale.}
  \label{fig:fiducial}
\end{figure}

Seven material samples (so-called fingers), made of a copper-beryllium alloy were used in this work. These samples were part of a plugin module attached to the CMS beampipe and used to maintain the beampipe's structural integrity upon changes in temperature.  Pictures of the plugin module with attached fingers, one of the finger samples, and an X-ray image of the the plugin module mounted in the beampipe are shown in Fig.~\ref{fig:xray}. The finger samples were replaced prior to luminosity running in 2012 since they were found to have been erroneously mounted and hung in the vacuum region. All material that has been part of the accelerator gets irradiate to some extent so safety procedures were necessary for the removal of these fingers. Radioactivity tests indicated that samples were safe for use in this study.

The sample length of an individual finger sample is around $8.5$~cm and the thickness is $\sim 0.8$~mm. The samples were located at a position of $18$~m along the colliding beam axis. The nominal polar angle acceptance range with respect to the CMS interaction point is between 179.936 and 179.943 degrees. A monopole produced within this angular range would stop inside the sample if produced with an energy below a certain punch-through energy, which is a function of its mass and charge. The Bethe formula for magnetic charges~\cite{Ahlen78} was used to simulate the energy loss of monopoles in small steps inside the sample material and estimate its range. The same approach was used as was employed in Ref.~\cite{SMPpheno11}. The results of these calculation are shown in Fig.~\ref{fig:fiducial}. The effective thickness of the samples is around 8 times larger than the actual thickness owing to the angle in which the samples lay. In addition, a monopole should have a minimum kinetic energy such that the $3.8$~T CMS magnetic field has a negligible effect on its trajectory. For masses above several hundred GeV this minimum energy is estimated to be of the order of 0.1~GeV. Given the limited range of samples currently available, the intention of this work is not to produce a limit on the production of monopoles. An analysis of the effects of the uncertainties of the size and positions of the samples is therefore not performed.


\section{Beampipe sample measurements}\label{sec:beampipe}

The sample holder used for the beampipe sample runs was a 50 cm long carbon-fibre hollow tube. The seven beampipe samples were wrapped into paper so that they would tightly fit into a slit on the tip of the sample holder. At the starting position of each run, the sample holder was inside the magnetometer with its tip  protruding on the side opposite to the holder arm. In this position, a beampipe  sample could be attached to the tip of the sample holder. Measurements were then performed in steps until the sample had passed entirely through the magnetometer. At the end position, nothing remained inside the magnetometer and the sample could be retrieved on the away side. Most measurements were performed with 48 steps to precisely map the magnetisation of the sample at each position. Each beampipe sample was measured at least twice, in 48 steps. Frequent empty holder measurements were made for background subtraction between beampipe samples measurements. Beampipe sample number 4 was measured six times, of which three had a different number of steps (96, 24 an 4).

\begin{figure}[tb]
  \begin{center}
    \includegraphics[width=0.7\linewidth]{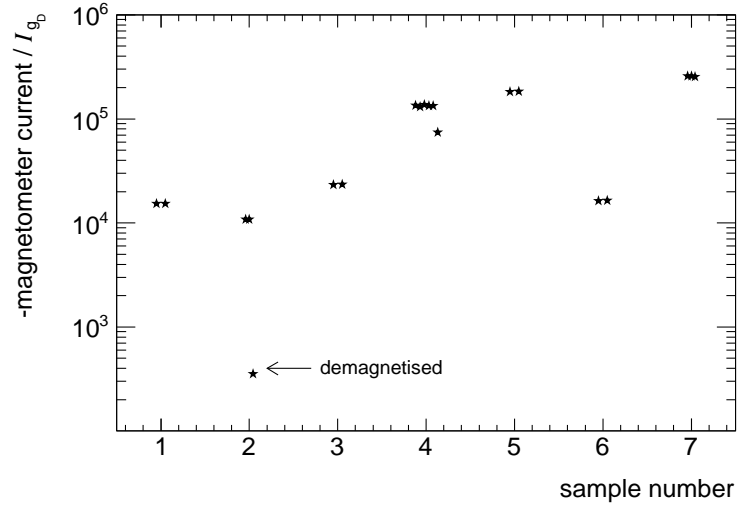}
  \end{center}
  \caption{Peak currents associated with different beampipe samples. }
  \label{fig:figpeak}
\end{figure}

 The magnetisation of the beampipe samples can be quantitatively studied by measuring the peak induced current, i.e. that value occurring when the beampipe samples are near the centre of the pick-up coil system. The peak currents are shown in Fig.~\ref{fig:figpeak}. As can be seen the currents are typically many orders of magnitude greater than that expected for a Dirac monopole. As mentioned in Section~\ref{sec:expmet}  offset jumps are expected for strongly magnetised samples. As a test to reduce the magnetisation, one sample (beampipe sample 2) was demagnetised with an oscillating magnetic field. This reduces the magnetisation by more than an order of magnitude, albeit at the possible risk of dislodging a trapped monopole. This risk
should, however, be minimal since the binding energy of a monopole in material
is thought to be large~\cite{Milton:2006cp}. Should offset jumps represent a limiting effect for further studies demagnetisation may be necessary.

\begin{figure}[tb]
  \begin{center}
    \includegraphics[width=0.7\linewidth]{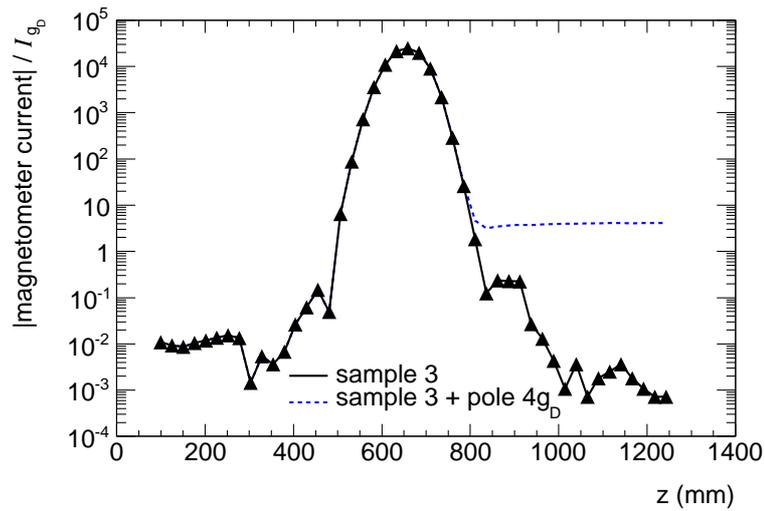}
  \end{center}
  \caption{The measured current from beampipe sample 3 as a function of $z$. Also shown (dashed line) is the signature which would be
  expected should the beampipe sample carry a magnetic charge of $4.14g_D$.}
  \label{fig:finger3}
\end{figure}

 The measurement of induced current for one run made with beampipe sample 3 is shown in Fig.~\ref{fig:finger3}, after subtracting the values obtained with an empty sample holder from a run performed immediately afterwards. The empty-holder subtraction corrects for the offset, and also for the very slight remnant magnetisation of the sample holder while it is still inside the magnetometer. The sum of the beampipe sample measurements and a solenoid measurement with pseudopole equivalent to $4.14g_D$ is shown as a dashed line superimposed to the beampipe sample measurements in Fig.~\ref{fig:finger3}, demonstrating that a sample with a trapped monopole would yield a very distinct (and reproducible) signature.

\begin{figure}[tb]
  \begin{center}
  \hspace{-0.0cm}
    \includegraphics[width=0.7\linewidth]{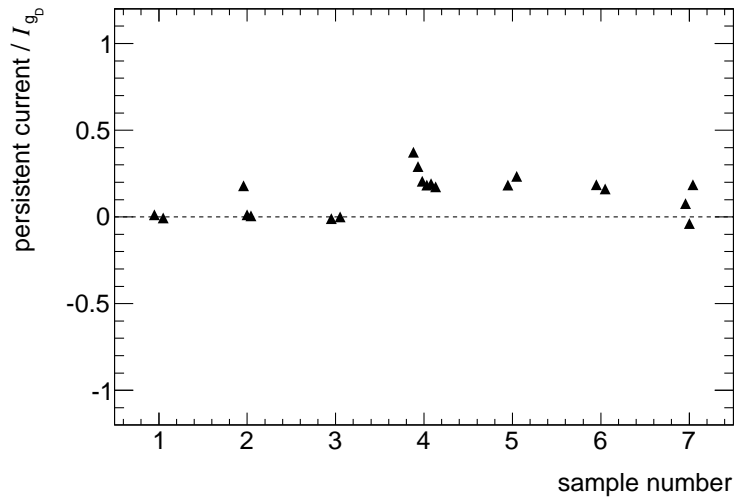}
  \end{center}

  \caption{Persistent current left by the beampipe samples. }
  \label{fig:fingers}
\end{figure}

The persistent current for all beampipe sample runs are shown in Fig.~\ref{fig:fingers}, after subtracting the value obtained from empty holder measurements. Values of persistent current range from 0 to $\sim 0.35I_{g_D}$ which is consistent with observations made with the rock samples (Tab.~\ref{tab:fakes}). When making the measurements it was observed that offset jumps of magnitude $\sim 0.15$~$g_D$ took place. Sample number 4 suffered several consecutive offset jumps which accounts for the higher currents observed for two of the measurements made with this sample giving the highest current readings.

\section{Summary and conclusions}
Monopoles and dyons with specific ranges of magnetic charge would be trapped in the beampipes of the LHC experiments and could be observed by measurements of the magnetic properties of the beampipe material. Using a SQUID-based magnetometer, a search has been made for stopped monopoles in obsolete accelerator material from the LHC. The main purpose of the work is to quantify the expected response of the magnetometer to monopoles, study how fake signals can arise and run through the protocols needed for a sample release from CERN. Several methods were used to calibrate the magnetometer and demonstrate its capability for detecting monopoles. The noise of the magnetometer was studied with magnetic rocks. The production kinematics of monopoles leading to stopping in the samples were also evaluated. With the limited range of available material samples, no evidence was found for monopoles with charges greater than around one third of the Dirac charge which represents the lowest value of magnetic charge which could be observed in this study.

\section{Acknowledgements}

As for any experimental undertaking at the LHC, this work would not have been possible without the aid of a large number of people.  The authors wish to thank Austin Ball, Stephane Bally and Martin
Gastal from the CMS collaboration for their help and support. We are also grateful to the TE, DGS, EN and GS departments at CERN. In particular we wish to thank Gerhard Schneider from the TE-VSC-LBV group, who
undertook the maintenance work which allowed the beampipe samples to become available. Abderrahim Errahhaoui performed essential radioactivity tests prior to the release of the samples. Machining work on the samples was carried out by Jacky Tonoli. Stephanie Krattinger performed the work which allowed the samples to be transported out of CERN.

This work was supported by a grant from the Swiss National Science Foundation.

%

%

\bibliographystyle{mystylem}
\bibliography{SQUIDFingers12}

\providecommand{\href}[2]{#2}\begingroup\raggedright\begin{thebibliography}{10}

\bibitem{Hooft1974}
G.~'t~Hooft, {\em {Magnetic Monopoles in Unified Gauge Theories}\/},
  \href{http://dx.doi.org/10.1016/0550-3213(74)90486-6}{Nucl. Phys. {\bf B79}
  (1974)  276}.

\bibitem{Preskill1984}
J.~Preskill, {\em {Magnetic Monopoles}\/},
Ann. Rev. Nucl. Part. Sci. {\bf 34} (1984)  461.

\bibitem{Dirac1931}
P.~A.~M. Dirac, {\em {Quantised Singularities in the Electromagnetic Field}\/},
   Proc. Roy. Soc. {\bf A 133} (1931)  60.

\bibitem{Dirac1948}
P.~A.~M. Dirac, {\em {The Theory of Magnetic Poles}\/},
  \href{http://dx.doi.org/10.1103/PhysRev.74.817}{Phys. Rev. {\bf 74} (1948)
  817}.

\bibitem{Fairbairn:2006gg}
M.~Fairbairn et al., {\em {Stable massive particles at colliders}\/},
  \href{http://dx.doi.org/10.1016/j.physrep.2006.10.002}{Phys. Rept. {\bf 438}
  (2007)  1},
\href{http://arxiv.org/abs/0611040}{{\tt arXiv:0611040 [hep-ph]}}.

\bibitem{Milton:2006cp}
K.~A. Milton, {\em {Theoretical and experimental status of magnetic
  monopoles}\/},
  \href{http://dx.doi.org/10.1088/0034-4885/69/6/R02}{Rept.Prog.Phys. {\bf 69}
  (2006)  1637--1712},
\href{http://arxiv.org/abs/hep-ex/0602040}{{\tt arXiv:hep-ex/0602040
  [hep-ex]}}.

\bibitem{PDG2010}
{Particle Data Group} Collaboration, {\em {Review of particle physics}\/},
\href{http://dx.doi.org/10.1088/0954-3899/37/7A/075021}{J. Phys. G {\bf 37}
  (2010)  075021}.

\bibitem{Schwinger:1966nj}
J.~S. Schwinger, {\em {Magnetic charge and quantum field theory}\/},
\href{http://dx.doi.org/10.1103/PhysRev.144.1087}{Phys. Rev. {\bf 144} (1966)
  1087}.

\bibitem{Zwanziger:1969by}
D.~Zwanziger, {\em {Exactly soluble nonrelativistic model of particles with
  both electric and magnetic charges}\/},
  \href{http://dx.doi.org/10.1103/PhysRev.176.1480}{Phys. Rev. {\bf 176} (1968)
   1480}.

\bibitem{Ahlen78}
S.~Ahlen, {\em {Stopping-power formula for magnetic monopoles}\/},  Phys. Rev.
  D {\bf 17} (1978)  229.

\bibitem{Ahlen80}
S.~Ahlen, {\em {Theoretical and experimental aspects of the energy loss of
  relativistic heavily ionizing particles}\/},  Rev. Mod. Phys. {\bf 52} (1980)
   121.

\bibitem{Moedal}
{MoEDAL} Collaboration, {\em {Technical Design Report of the Moedal
  Experiment}\/},  CERN-LHCC-2009-006 ; MOEDAL-TDR-001 (2009)  .

\bibitem{Pinfold:2009zz}
J.~L. Pinfold, {\em {Searching for the magnetic monopole and other highly
  ionizing particles at accelerators using nuclear track detectors}\/},
\href{http://dx.doi.org/10.1016/j.radmeas.2009.10.062}{Radiat. Meas. {\bf 44}
  (2009)  834}.

\bibitem{:1999fr}
{\em {ATLAS: Detector and physics performance technical design report. Volume
  2,}\/},
CERN-LHCC-99-15, ATLAS-TDR-15 (1999)  .

\bibitem{Aad:2011mb}
{ATLAS Collaboration} Collaboration, G.~Aad et al., {\em {Search for Massive
  Long-lived Highly Ionising Particles with the ATLAS Detector at the LHC}\/},
  Phys.Lett. {\bf B698} (2011)  353--370,
\href{http://arxiv.org/abs/1102.0459}{{\tt arXiv:1102.0459 [hep-ex]}}.

\bibitem{ATLASmonoconf}
{\em {Search for magnetic monopoles in $\sqrt{s}$=7 TeV pp collisions with the
  ATLAS detector,}\/},
ATLAS-CONF-2012-062 (2012)  .

\bibitem{SMPpheno11}
A.~De~Roeck, A.~Katre, P.~Mermod, D.~Milstead, and T.~Sloan, {\em {Sensitivity
  of LHC experiments to exotic highly ionising particles}\/},  Eur. Phys. J.
  {\bf C72} (2012)  1985,
\href{http://arxiv.org/abs/1112.2999}{{\tt arXiv:1112.2999 [hep-ph]}}.

\bibitem{Alvarez:1970zu}
L.~Alvarez, P.~Eberhard, R.~Ross, and R.~Watt, {\em {Search for magnetic
  monopoles in the lunar sample}\/},
Science {\bf 167} (1970)  701--703.

\bibitem{Alvarez:1971zt}
L.~Alvarez, M.~Antuna, R.~Byrns, P.~Eberhard, R.~Gilmer, et al., {\em {A
  magnetic monopole detector utilizing superconducting elements}\/},
\href{http://dx.doi.org/10.1063/1.1685086}{Rev.Sci.Instrum. {\bf 42} (1971)
  326--330}.

\bibitem{TEVATRONSQUID2000}
G.~R. Kalbfleisch, K.~A. Milton, M.~G. Strauss, L.~Gamberg, E.~H. Smith, and
  W.~Luo, {\em {Improved Experimental Limits on the Production of Magnetic
  Monopoles}\/},  \href{http://dx.doi.org/10.1103/PhysRevLett.85.5292}{Phys.
  Rev. Lett. {\bf 85} (2000)  5292}.

\bibitem{TEVATRONSQUID2004}
G.~R. Kalbfleisch, W.~Luo, K.~A. Milton, E.~H. Smith, and M.~G. Strauss, {\em
  {Limits on production of magnetic monopoles utilizing samples from the D0 and
  CDF detectors at the Tevatron}\/},
  \href{http://dx.doi.org/10.1103/PhysRevD.69.052002}{Phys. Rev. {\bf D69}
  (2004)  052002},
\href{http://arxiv.org/abs/0306045}{{\tt arXiv:0306045 [hep-ex]}}.

\bibitem{Aktas:2004qd}
{H1 Collaboration} Collaboration, A.~Aktas et al., {\em {A Direct search for
  stable magnetic monopoles produced in positron-proton collisions at HERA}\/},
   \href{http://dx.doi.org/10.1140/epjc/s2005-02201-6}{Eur.Phys.J. {\bf C41}
  (2005)  133--141},
\href{http://arxiv.org/abs/hep-ex/0501039}{{\tt arXiv:hep-ex/0501039
  [hep-ex]}}.

\bibitem{2g}
 2g Enterprises homepage, \url{http://www.2genterprises.com}, accessed June
  2012.

\bibitem{Maxwell}
 Maxwell SV homepage, \url{http://www.ansoft.co.kr/html/dow/maxwell.php},
  accessed June 2012.

\bibitem{parker}
R.~L. Parker, {\em {Calibration of the pass-through magnetometer - I.
  Theory}\/},
Geophys. J. Int. {\bf 142} (2000)  371--383.

\bibitem{jackson}
M.~Jackson et al., {\em {Deconvolution of u channel magnetometer data:
  Experimental study of accuracy, resolution, and stability of different
  inversion methods}\/},
Geochem. Geophys. Geosyst. {\bf 11} (2010)  Q07Y10.

\bibitem{weeks}
R.~Weeks et al., {\em {Improvements in long-core measurement techniques:
  applications in palaeomagnetism and palaeoceanography}\/},
Geophys. J. Int. {\bf 114} (1993)  651--662.

\bibitem{chen}
H.~Chen, {\em {Note on the Magnetic Pole}\/},
Am. Jour. Phys. {\bf 33} (1965)  563--565.

\bibitem{Levy}
L.-P. Levy, {\em {Magnetism and Superconductivity}\/},  Springer (2004)  .

\bibitem{R1andR2}
 Basalt volcanic rock samples R1 (PRR-6165) and R2 (PRR-6166),
  \url{http://bprc.osu.edu/rr/}, accessed June 2012.

\end{thebibliography}\endgroup


\providecommand{\href}[2]{#2}\begingroup\raggedright\endgroup

\end{document}